\documentclass[12pt,a4paper]{iopart}
\usepackage{iopams}
\usepackage{graphicx}
\usepackage[usenames]{color}
\usepackage[normalem]{ulem}

\begin{document}

\title[Vortical and fundamental solitons in dipolar BEC]{Vortical and
fundamental solitons in dipolar Bose-Einstein condensates trapped in
isotropic and anisotropic nonlinear potentials}
\author{R. Kishor Kumar$^{1}$, P. Muruganandam$^{1}$ and B. A. Malomed$^{2}$}
\address{$^1$School of Physics, Bharathidasan University, Palkalaiperur Campus,
Tiruchirappalli 620024, Tamilnadu, India\\
$^2$ Department of Physical Electronics, School of Electrical
Engineering, Faculty of Engineering, Tel Aviv University, Tel Aviv
69978, Israel}

\begin{abstract}
We predict the existence of stable fundamental and vortical bright
solitons in dipolar Bose-Einstein condensates (BECs) with repulsive
dipole-dipole interactions (DDI). The condensate is trapped in the
2D plane with the help of the repulsive contact interactions whose
local strength grows $\sim r^{4}$ from the center to periphery,
while dipoles are oriented perpendicular to the self-trapping plane.
The confinement in the perpendicular direction is provided by the
usual harmonic-oscillator potential. The objective is to extend the
recently induced concept of the self-trapping of bright solitons and
solitary vortices in the \textit{pseudopotential}, which is induced
by the repulsive local nonlinearity with the strength growing from
the center to periphery, to the case when the trapping mechanism
competes with the long-range repulsive DDI. Another objective is to
extend the analysis for elliptic vortices and solitons in an
anisotropic nonlinear pseudopotential. Using the variational
approximation (VA) and numerical simulations, we construct families
of self-trapped modes with vorticities $\ell =0$ (fundamental
solitons), $\ell =1$, and $\ell =2$. The fundamental solitons and
vortices with $\ell =1$ exist up to respective critical values of
the eccentricity of the anisotropic pseudopotential, being stable in
the entire existence regions. The vortices with $\ell =2$ are stable
solely in the isotropic model.
\end{abstract}

\maketitle

\section{Introduction}

Self-trapping of solitary modes, both fundamental and vortical ones, is one
of basic topics in the current research work in the fields of nonlinear
optics, matter-wave dynamics in quantum gases, and related areas \cite%
{book,review,big-review,Yang,Peli,RMP}. Usually, the self-trapping is
provided by the self-focusing/self-attractive nonlinearity. Self-repulsion,
if acting in the combination with periodic potentials, gives rise to
localized modes of the bandgap type \cite{Yang,Peli}. Spatial modulation of
the local nonlinearity may also help to support solitons \cite{RMP}, but it
was assumed that the nonlinearity should be self-focusing on average in that
case. Only very recently it has been demonstrated that purely repulsive
nonlinear interactions may support bright solitons and solitary vortices, if
the strength of the local cubic self-defocusing term grows with the distance
from the center ($r$) faster than $r^{D}$, where $D$ is the spatial
dimension \cite{Barcelona}. In addition to straightforward realizations in
optics, where the necessary spatial modulation of the local nonlinearity can
be created by means of inhomogeneous doping \cite{Kip}, this setting may be
realized in Bose and Fermi gases \cite{Luca}, making use of the Feshbach
resonance (FR) controlled by spatially nonuniform magnetic \cite%
{Feshbach-magn}, optical \cite{Feshbach-opt}, or dc electric \cite%
{Feshbach-dc} fields. An extension of the setting for the model of an
optical medium with the \emph{nonlocal} self-defocusing thermal nonlinearity
\cite{thermal} has demonstrated that stable one-dimensional (1D) fundamental
and dipole-mode solitons can be supported in this setting too, provided that
the strength of the heating term grows with the coordinate faster than $|x|$
\cite{Yingji}.

A particular topic in the studies of Bose-Einstein condensates (BECs), which
has drawn a great deal of attention, is dealing with quantum gases of
dipolar atoms and molecules \cite{RPP}. Soliton modes were analyzed in
models of dipolar BECs in various continuous \cite{DD-sol} and discrete \cite%
{sol-discr} settings (similar one-dimensional solitons were also predicted

in the model of the dipolar Tonks-Girardeau gas \cite{TG}). In this context,
stable two-dimensional (2D) solitons \cite{2D-DD-sol} and solitary vortices
\cite{Tikh2008} were predicted in the isotropic form, assuming an inverted
sign of the dipole-dipole interaction (DDI). Anisotropic fundamental 2D
solitons were also found in a model based on the natural DDI sign \cite%
{Malomed2008,Patrick}. Numerical results for gap solitons in the dipolar BEC
trapped in the 3D optical lattice have been reported too \cite{3D}.

The objective of the present work is to consider the possibility of trapping
the dipolar condensate in the plane of $\left( x,y\right) $, with the
repulsive DDI between dipole moments polarized in the perpendicular
direction, along axis $z$. The trapping is provided by the repulsive contact
interaction, whose strength is assumed to grow $\sim r^{4}$ in the
horizontal plane, and by the usual linear harmonic-oscillator (HO) potential
applied in the vertical direction.

If the DDI is provided by atomic or molecular magnetic moments, the assumed
spatial modulation of the contact self-repulsion should be imposed by the FR
controlled by the optical or electric field, as the external magnetic field
serves to polarize the magnetic moments. The power-law growth of the
repulsion strength does not imply that the respective control field must
grow similarly. Instead, it is enough to provide a weak transverse
modulation of the field, which gradually reduces the local detuning of the
FR with the increase of $r$, on the characteristic scale of $r\sim 100~%
\mathrm{\mu }$m \cite{Barcelona}. Such a weak modulation can be
easily implemented by making the laser beam slightly focusing or
defocusing, or applying the dc electric field induced by the
capacitor with slightly non-flat electrodes. If, on the other hand,
the DDI is provided by molecular electric dipoles, they must be
polarized by means of an external electrostatic field, while a
magnetic or optical field may be used to control the local detuning
of the FR. In particular, a weakly nonuniform magnetic field may be
applied by means of a tapered solenoid \cite{tapering} or
appropriately designed magnetic lattice \cite{magnetic-lattice}.

In this work, we consider both isotropic and anisotropic shapes of the
effective nonlinear pseudopotential acting in the $\left( x,y\right) $
plane. It is relevant to mention that the existence and stability of
elliptic vortices in anisotropic HO traps was previously studied for the
self-repulsive \cite{ell-rep} and attractive \cite{ell-attr} signs of the
local cubic nonlinearity, but anisotropic nonlinear pseudopotentials were
not considered before in any form, to the best of our knowledge.

The analysis is based on the variational approximation (VA) and direct
simulations in imaginary and real time. The VA provides an initial
approximation for the imaginary-time integration (it is known that this
method is relevant for constructing both fundamental and vortical stationary
modes \cite{Im}), while the stability of the stationary modes is tested by
means of the real-time propagation. We aim to construct vortical modes with
topological charges $\ell =1$ and $2$, as well as fundamental solitons with $%
\ell =0$.

The paper is structured as follows. The model and VA are introduced in
Section II. The results for the vortices and fundamental solitons are
reported in Sections III and IV, respectively. In particular, we find that
the modes with $\ell =1$ and $0$ are stable in their full existence domains,
up to the largest eccentricity which admits the existence of the trapped
modes in the anisotropic nonlinear pseudopotential, while the double
vortices with $\ell =2$ are stable only in the isotropic pseudopotential.
The paper is concluded by Section V.

\section{The mean-field description}

\label{sec:frame}

In the mean-field approximation, a dipolar BEC with $N$ atoms of mass $m$ at
zero temperature is described by the Gross-Pitaevskii equation (GPE),

\begin{eqnarray}
i\frac{\partial \psi (\mathbf{r},t)}{\partial t}=\left[
-\frac{1}{2}\nabla
^{2}+\frac{1}{2}z^{2}+gV_{\mathrm{NL}}(\mathbf{r})|\psi (\mathbf{r}%
,t)|^{2}\right.   \nonumber \\
\left. +g_{\mathrm{d}}\int U_{\mathrm{DD}}(\mathbf{r-r^{\prime }})|\psi (%
\mathbf{r^{\prime }},t)|^{2}d\mathbf{r}^{\prime }\right] \psi (\mathbf{r},t),
\label{eqn:dgpe}
\end{eqnarray}%
where $\psi (\mathbf{r},t)$ is the mean-field wave function subject
to the normalization condition,
\begin{equation}
\int |\psi (\mathbf{r},t)|^{2}d\mathbf{r}=1,  \label{N}
\end{equation}%
and the kernel of the DDI is $U_{\mathrm{DD}}({\mathbf{r}})=\left( 1-3\cos
^{2}\theta \right) r^{-3}$, where $\theta $ is the angle between ${\mathbf{r}%
}$ and the direction of the polarization of the dipoles, $z$. In this
direction, the condensate is confined by the HO potential with trapping
frequency $\omega _{z}$. Accordingly, in equation (\ref{eqn:dgpe}) lengths
are measured in units of the respective HO length, $l_{z}\equiv \sqrt{\hbar
/m\omega _{z}}$, frequency in units of $\omega _{z}$, and time $t$ in units
of $\omega _{z}^{-1}$, the contact-interaction and DDI strengths being,
respectively, $g=4\pi a_{s}N/l_{z}$ and $g_{\mathrm{d}}=Nd^{2}m/(\hbar
^{2}l_{z})$, where $a_{s}$ is the scattering length of atomic collisions,
and $d$ is the magnetic moment.

As said above, the repulsive contact interactions are subject to the
modulation in the $\left( x,y\right) $ plane, with local strength growing
from the center to periphery,
\begin{equation}
V_{\mathrm{NL}}=(1+\epsilon _{1}x^{2}+\epsilon _{2}y^{2})^{2},  \label{V}
\end{equation}%
which corresponds to the action of an effective nonlinear quartic potential.
As said above, in the 2D space we need to introduce the
nonlinearity-strength-modulation function growing faster than $r^{D}\equiv
r^{2}$, therefore we here adopt the quartic radial profile. Generally, the
pseudopotential corresponding to the azimuthal modulation profile in (\ref{V}%
) is anisotropic, with eccentricity%
\begin{equation}
\delta \equiv \frac{\epsilon _{2}-\epsilon _{1}}{\epsilon _{1}+\epsilon _{2}}%
.  \label{delta}
\end{equation}

\section{Vortex solitons}

\label{sec:vs}

\subsection{The energy minimization}

In terms of cylindrical coordinates $z$, $\rho $, and $\phi $, stationary
solutions to Eq. (\ref{eqn:dgpe}) for vortices with integer topological
charge $\ell \geq 0$ are looked for in the usual form,
\begin{equation}
\psi _{\ell }=\exp \left( -i\mu t+i\ell \phi \right) \rho ^{\ell }U_{\ell
}\left( \rho ,z,\phi \right) ,  \label{U}
\end{equation}%
where function $U_{\ell }$ remains finite at $\rho =0$, and decays as $\rho
^{\ell }\left\vert U_{\ell }\right\vert \approx \sqrt{\mu /g}\left( \epsilon
_{1}x^{2}+\epsilon _{2}y^{2}\right) ^{-1}$ at $\rho \rightarrow \infty $,
according to the Thomas-Fermi approximation \cite{Barcelona}. The dependence
on angular coordinate $\phi $ in this function is present in the case of
anisotropic modulation pattern (\ref{V}). In the latter case, $U_{\ell }$ is
complex, while it is real for the isotropic setting.

The energy functional corresponding to GPE (\ref{eqn:dgpe}) is
\begin{eqnarray}
E=\frac{1}{2}\int \left[ \left\vert \nabla \psi \right\vert
^{2}+z^{2}\left\vert \psi \right\vert ^{2}+gV_{\mathrm{NL}}(\mathbf{r})|\psi
(\mathbf{r},t)|^{4}\right.   \nonumber \\
+g_{\mathrm{d}}\int \left. U_{\mathrm{DD}}(\mathbf{r-r^{\prime }})|\psi (%
\mathbf{r},t)|^{2}|\psi (\mathbf{r^{\prime }},t)|^{2}d\mathbf{r}^{\prime }%
\right] d\mathbf{r}.  \label{E}
\end{eqnarray}%
To predict vortex-soliton modes (\ref{U}) by means of the VA, we
calculate the value of functional (\ref{eqn:dgpe}), using expression
(\ref{U}) with a
real Gaussian ansatz for $U_{\ell }\left( \rho ,z\right) $, cf. Ref. \cite%
{Tikh2008}:
\begin{equation}
U_{\ell }(\rho ,z)=A_{\ell }\exp \left[ -\frac{1}{2}\left( \alpha \rho
^{2}+\gamma z^{2}\right) \right] ,  \label{eqn:vs-ansatz}
\end{equation}%
where $\alpha $ and $\rho $ are free variational parameters, and the
amplitude is determined by normalization condition (\ref{N}):%
\begin{equation}
A_{\ell }=\pi ^{-3/4}\left( \ell !\right) ^{-1/2}\sqrt{\alpha ^{\ell +1}}%
\gamma ^{1/4}.  \label{A}
\end{equation}%
Note that, even in the case of the anisotropic modulation profile (\ref{V}),
the isotropic form is adopted for ansatz (\ref{eqn:vs-ansatz}), as otherwise
the application of the VA to vortices is too cumbersome. The results for $%
\ell =1$ and $2$ are%
\begin{eqnarray}
E_{1}=\alpha +\frac{1}{4}\left( \gamma +\frac{1}{\gamma }\right) +\frac{1}{2}%
\sqrt{\frac{\gamma }{2\pi }}\left[ \frac{g}{32\pi }\right.   \nonumber \\
\left. \times \frac{9\epsilon _{1}^{2}+9\epsilon _{2}^{2}+12\epsilon
_{2}\alpha +8\alpha ^{2}+6\epsilon _{1}(\epsilon _{2}+2\alpha )}{\alpha }+%
\frac{g_{\mathrm{d}}}{3}\alpha f_{1}(\kappa )\right] ,  \label{1}
\end{eqnarray}%
\begin{eqnarray}
E_{2}=\frac{3}{2}\alpha +\frac{1}{4}\left( \gamma +\frac{1}{\gamma }\right) +%
\frac{3}{8}\sqrt{\frac{\gamma }{2\pi }}\left[ \frac{g}{64\pi }\right.
\nonumber \\
\left. \times {\frac{15(3\epsilon _{1}^{2}+2\epsilon _{1}\epsilon
_{2}+3\epsilon _{2}^{2})+40(\epsilon _{1}+\epsilon _{2})\alpha +16\alpha ^{2}%
}{\alpha }}+\frac{g_{\mathrm{d}}}{3}\alpha f_{2}(\kappa )\right] ,  \label{2}
\end{eqnarray}%
where $\kappa \equiv \sqrt{\gamma /\alpha }$ is the aspect ratio of ansatz (%
\ref{eqn:vs-ansatz}), and the following \textit{ad hoc} functions are
defined:
\begin{eqnarray}
&&f_{1}(\kappa )\equiv -1+3\int_{0}^{1}R(\kappa ,x)\left[ 1+Q^{2}(\kappa ,x)%
\right] dx, \\
&&f_{2}(\kappa )\equiv -1+3\int_{0}^{1}R(\kappa ,x)\left[ 1+\frac{2}{3}%
Q^{2}(\kappa ,x)+Q^{4}(\kappa ,x)\right] dx, \\
&&R(\kappa ,x)\equiv \frac{(\kappa x)^{2}}{(\kappa x)^{2}+(1-x^{2})}%
,~Q(\kappa ,x)\equiv \frac{1-x^{2}}{(\kappa x)^{2}+(1-x^{2})}.
\end{eqnarray}

\subsection{Vortices with $\ell =1$}

\begin{figure}[th]
\begin{center}
\includegraphics[width=0.9\columnwidth,clip]{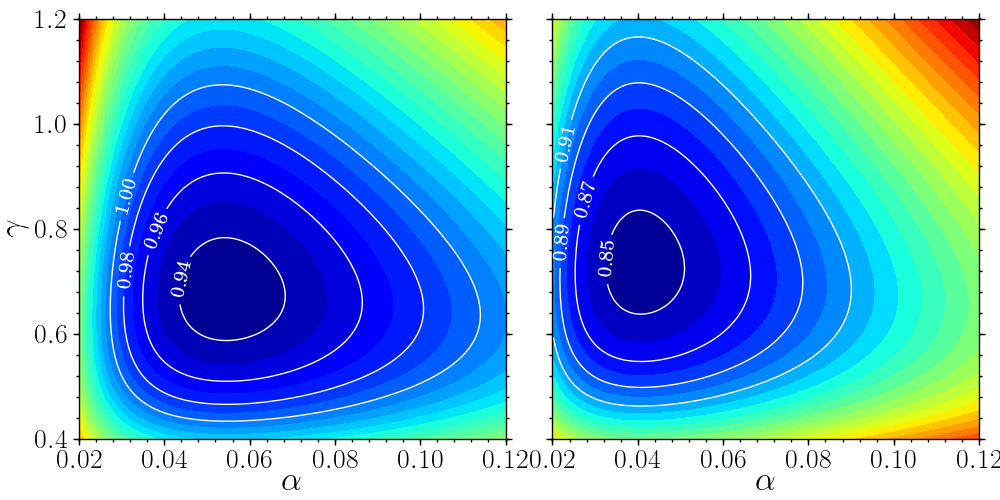}
\end{center}
\caption{(Color online) GPE energy functional (\protect\ref{E}) for the
vortex-soliton ansatz (\protect\ref{eqn:vs-ansatz}) with $\ell =1$ as a
function of variational parameters $\protect\alpha $ and $\protect\gamma $.
The interaction strengths are $g=20$, $g_{d}=30$, and parameters of the
quartic isotropic and anisotropic nonlinear potential (\protect\ref{V}) are $%
\protect\epsilon _{1}=\protect\epsilon _{2}=0.1$ in (a), and $\protect%
\epsilon _{1}=0.05,\protect\epsilon _{2}=0.1$ in (b).}
\label{fig1}
\end{figure}

As seen from contour plots of the energy displayed in figures \ref{fig1}(a)
and \ref{fig1}(b) for $\ell =1$, in both the isotropic and anisotropic
versions of the model the energy, considered as a function of the
variational parameters, has a well-defined minimum, which should correspond
to a (presumably stable) stationary vortex soliton. Values of the parameters
corresponding to this solution were found from a numerical solution of the
equations defining the energy minimum:
\begin{equation}
\frac{\partial E}{\partial \alpha }=\frac{\partial E}{\partial \gamma }=0.
\label{dddd}
\end{equation}

Next, the vortex-soliton profiles predicted by the VA were used as initial
conditions to generate their numerically exact counterparts by means of the
imaginary-time integration method, in the isotropic and anisotropic models
alike. Numerical simulations, in the imaginary and real time alike, were
carried out using a method combining the split-step Crank-Nicolson algorithm
and fast Fourier transform~\cite{CPC}.

\begin{figure}[th]
\begin{center}
\includegraphics[width=0.9\columnwidth,clip]{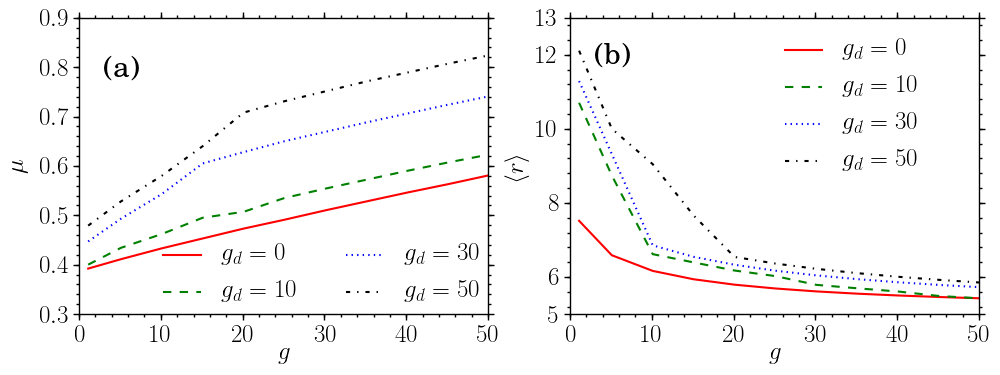}
\end{center}
\caption{(Color online) The numerically calculated chemical potential (a),
and the rms radius (b), as functions of the contact-interaction strength, at
fixed values of the DDI strength, for vortex solitons with topological
charge $\ell =1$ trapped in the isotropic nonlinear pseudopotential with $%
\protect\epsilon _{1}=\protect\epsilon _{2}=0.1$.}
\label{fig2}
\end{figure}

\begin{figure}[th]
\begin{center}
\includegraphics[width=0.9\columnwidth,clip]{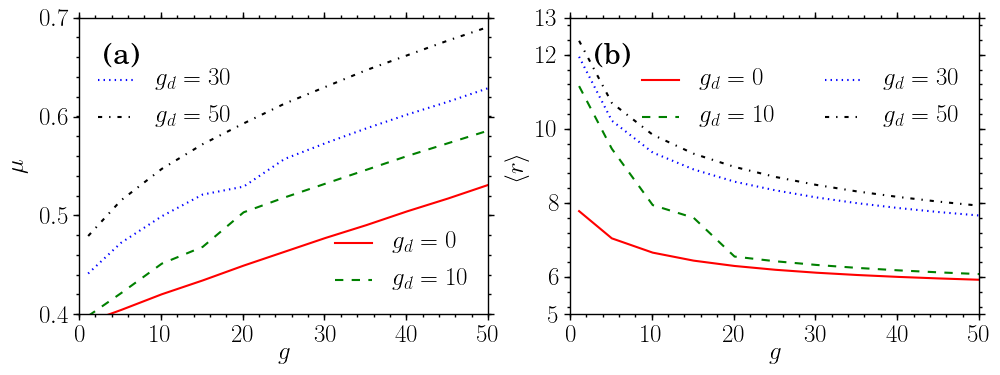}
\end{center}
\caption{(Color online) The same as in figure \protect\ref{fig2}, but for
elliptic vortices trapped in the anisotropic pseudopotential with $\protect%
\epsilon _{1}=0.05$, $\protect\epsilon _{2}=0.1$.}
\label{fig3}
\end{figure}

The results demonstrate that a difference $\lesssim 10\%$ between the
VA-predicted and numerically generated vortex profiles. The numerically
constructed families of the vortex solitons in the isotropic and anisotropic
systems are illustrated, respectively, in figures \ref{fig2} and \ref{fig3}
by plots showing the dependence of the chemical potential [see equation (\ref%
{U})] and root-mean-square (rms) radius of the vortices on strength $g$ of
the local repulsion, at several constant values of the long-range repulsion
strength, $g_{\mathrm{d}}$, including $g_{\mathrm{d}}=0$, for the sake of
the comparison with the usual model which does not include the DDI. The
decrease of the radius with the increase of $g$, observed in figures \ref%
{fig2}(b) and \ref{fig3}(b), is a natural property of solitons supported by
the growing repulsive nonlinearity \cite{Barcelona}. It is natural too that
the radius increases with the strength of the additional repulsive DDI,
which is seen in the same plots.

\begin{figure}[th]
\begin{center}
\includegraphics[width=0.6\columnwidth,clip]{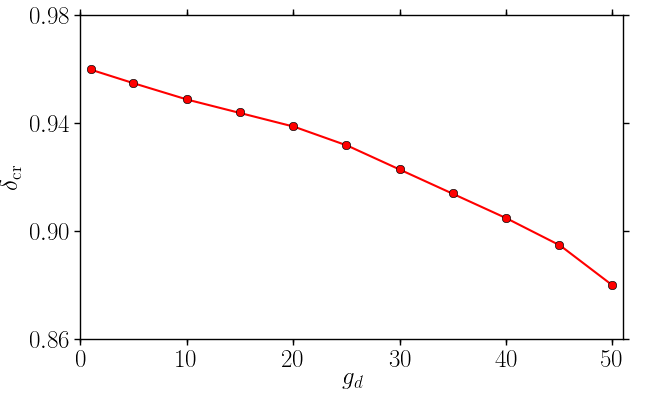}
\end{center}
\caption{(Color online) The critical eccentricity for the existence of
elliptic vortices with $\ell =1$, trapped in the anisotropic pseudopotential
with $g=20$, as a function of the strength of the dipole-dipole repulsion.}
\label{fig6}
\end{figure}

In the anisotropic model, it has been found that the vortex solitons with $%
\ell =1$ exist when eccentricity (\ref{delta}) takes values below a certain
critical value, $\delta _{\mathrm{cr}}$, which may be quite close to $1$,
i.e., the critical shape, beyond which the vortex does not exist, may be
strongly elongated. In particular,
\begin{equation}
\left( \delta _{\mathrm{cr}}\right) _{\ell =1}=0.923  \label{crit}
\end{equation}%
was found for the elliptic vortex with $g=20$ and $g_{\mathrm{d}}=30$. A
systematic dependence of the critical eccentricity on the DDI strength is
presented by the plot in figure \ref{fig6}. The stronger isotropic DDI
naturally tends to suppress the ellipticity, which explains the gradual
decrease of $\delta _{\mathrm{cr}}$ with the increase of $g_{\mathrm{d}}$.

Then, the stability of the so generated modes was tested by means of
simulations of their perturbed evolution in real time. The conclusion is
that all the solitary vortices with $\ell =1$ are stable in the isotropic
and anisotropic settings alike, see examples of the evolution in figures \ref%
{fig4} and \ref{fig5}, respectively (in fact, these figures also
display typical shapes of the isotropic and anisotropic vortices).
In this connection, it is relevant to mention that the increase of
the chemical
potential with $g$, which is observed in figures \ref{fig2}(a) and \ref{fig3}%
(a), implies that the vortex families satisfy the "anti-Vakhitov-Kolokolov"
criterion, which plays the role of the necessary stability condition for
localized modes supported by repulsive nonlinearities~\cite{anti}.
\begin{figure}[th]
\begin{center}
\includegraphics[width=0.9\columnwidth,clip]{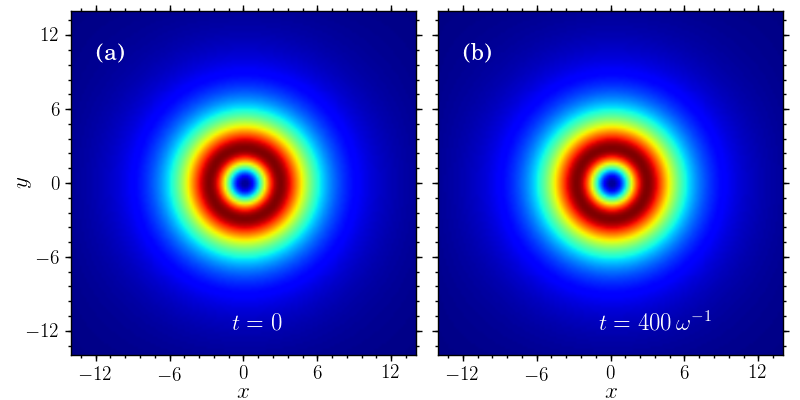}
\end{center}
\caption{(Color online) An example of stable evolution of the vortex soliton
with $\ell =1$ in the isotropic system. Shown are the initial state,
obtained by means of the imaginary-time propagation, starting from Gaussian
ansatz (\protect\ref{eqn:vs-ansatz}) with $\protect\alpha =0.05412$ and $%
\protect\gamma =0.6798$, and the final result of subsequent real-time
simulations. The parameters are $g=20$, $g_{d}=30$, and $\protect\epsilon %
_{1}=\protect\epsilon _{2}=0.1$. }
\label{fig4}
\end{figure}
\begin{figure}[th]
\begin{center}
\includegraphics[width=0.9\columnwidth,clip]{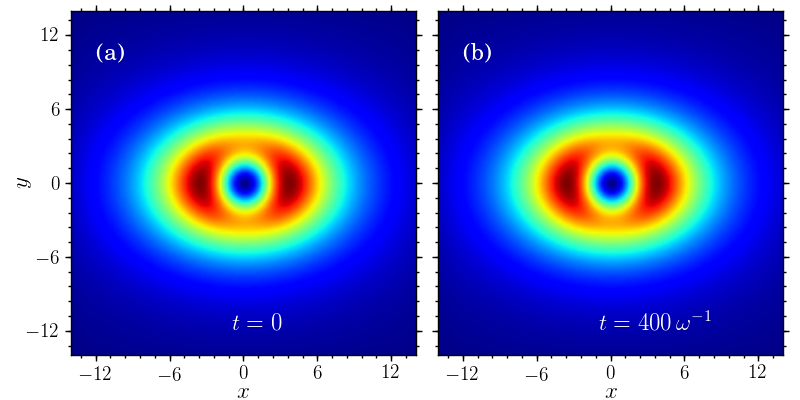}
\end{center}
\caption{(Color online) The same as in figure \protect\ref{fig4}, but for
the elliptic vortex in the anisotropic system with $\protect\epsilon %
_{1}=0.05$ and $\protect\epsilon _{2}=0.1$. The initial state is produced by
the imaginary-time propagation, starting from Gaussian ansatz (\protect\ref%
{eqn:vs-ansatz}) with $\protect\alpha =0.0403$ and $\protect\gamma =0.7316$.}
\label{fig5}
\end{figure}

\subsection{Vortices with $\ell =2$}

Vortex solitons with the double topological charge are also stable in the
presence of the isotropic nonlinear pseudopotential. As shown in figure \ref%
{fig7}, in the course of long evolution, a perturbation initially added to
the isotropic vortex remains trapped in it and causes small persistent
irregular perturbations, but does not lead to irreversible fragmentation of
the vortex (the evolution picture does not alter, at least, up to $%
t=2500\,\omega ^{-1}$, which was the full simulation time).
\begin{figure}[th]
\begin{center}
\includegraphics[width=0.9\columnwidth,clip]{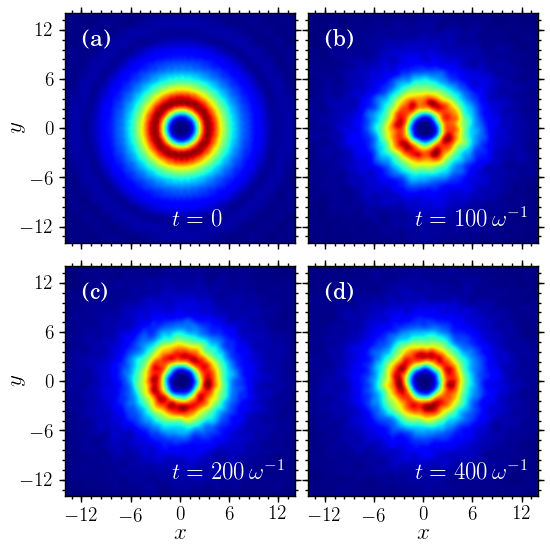}
\end{center}
\caption{(Color online) The persistent perturbed evolution of the vortex
mode with $\ell =2$ under the action of the interactions with strengths $g=20
$, $g_{d}=30$, and the isotropic quartic nonlinear pseudopotential (\protect
\ref{V}) with $\protect\epsilon _{1}=\protect\epsilon _{2}=0.1$.}
\label{fig7}
\end{figure}

\begin{figure}[th]
\begin{center}
\includegraphics[width=0.9\columnwidth,clip]{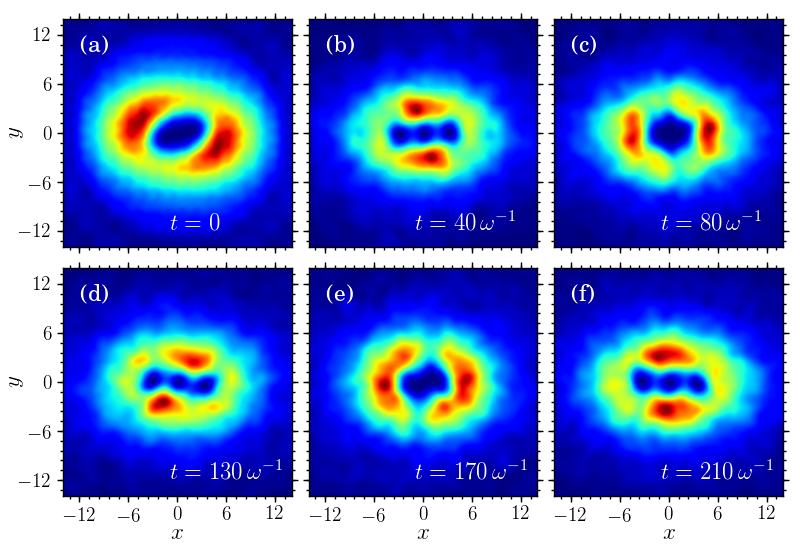}
\end{center}
\caption{(Color online) The evolution of the elliptic vortex soliton with $%
\ell =2$ in the dipolar BEC for $g=20$, $g_{d}=30$, trapped by the
anisotropic quartic nonlinear potential (\protect\ref{V}), with $\protect%
\epsilon _{1}=0.05$, $\protect\epsilon _{2}=0.1$.}
\label{fig8}
\end{figure}

On the other hand, under the anisotropic quartic pseudopotential, even with
a small eccentricity, the double vortex in subject to a splitting
instability, as can be seen in figure~\ref{fig8}.
In fact, the vortical structure is not completely destroyed in this case,
but rather splits and recovers periodically. We followed this evolution up
to $t=3000\,\omega ^{-1}$, concluding that the periodic cycles of the
splitting and recombination represent a robust dynamical regime. It
resembles the known scenario for the evolution of 2D vortices with $\ell =1$
in the condensate with the uniform \emph{attractive }contact nonlinearity,
trapped in the linear \emph{isotropic} HO potential, at values of the
nonlinearity strength which are intermediate between regions of the full
stability and destructive instability (see, e.g., Ref. \cite{DumDum}).

\begin{figure}[th]
\begin{center}
\includegraphics[width=0.6\columnwidth,clip]{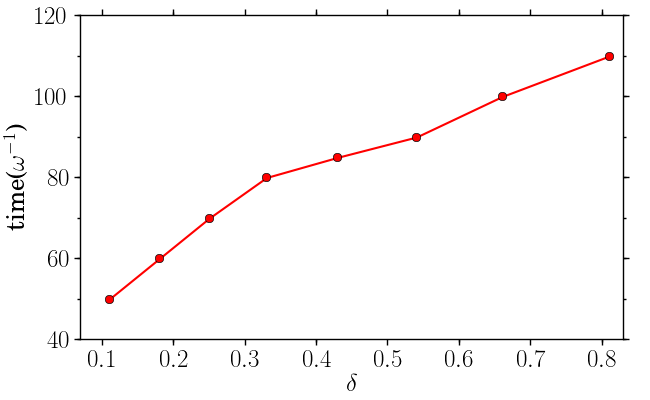}
\end{center}
\caption{(Color online) The period of the splitting-recombination regime (in
units of  $\protect\omega ^{-1}$) for vortices with $\ell =2$, versus the
eccentricity ($\protect\delta $).}
\label{fig9}
\end{figure}

On the basis of systematic simulations, we have concluded that the periodic
splitting-recombination regime persists for parameters at which stationary
elliptic vortices with $\ell =2$ can be found in the anisotropic
pseudopotential. The corresponding period is shown, as a function of
eccentricity (\ref{delta}), in figure \ref{fig9}. Naturally, the period
decreases with $\delta $, as the transition to the isotropic setting
stabilizes the double vortices.

\section{Fundamental solitons ($\ell =0$)}

\label{sec:fs}

\begin{figure}[th]
\begin{center}
\includegraphics[width=\columnwidth,clip]{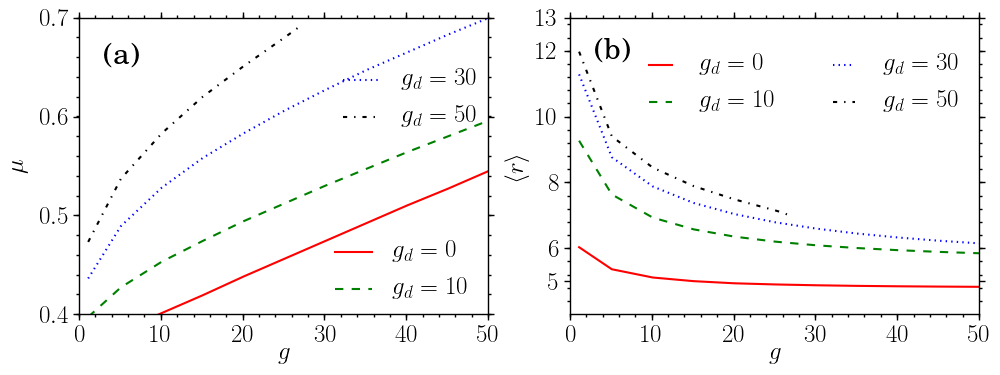}
\end{center}
\caption{(Color online) The same as in figure \protect\ref{fig2}, but for
the fundamental solitons ($\ell =0$).}
\label{fig10}
\end{figure}

\begin{figure}[th]
\begin{center}
\includegraphics[width=\columnwidth,clip]{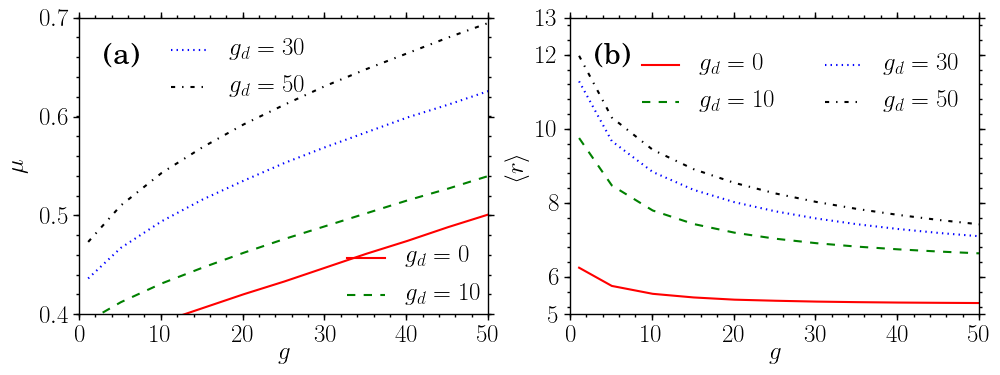}
\end{center}
\caption{(Color online) The same as in figure \protect\ref{fig3}, but for
elliptic fundamental solitons ($\ell =0$).}
\label{fig11}
\end{figure}

The VA for the fundamental solitons is based on equations (\ref%
{eqn:vs-ansatz})-(\ref{dddd}) with $\ell =0$. As it was done above for
vortices, numerical solutions have been produced by means of the
imaginary-time integration method, starting from the VA prediction. Similar
to the situation for the vortical modes, the difference between the
VA-predicted and numerically generated profiles of the fundamental solitons
is $\sim 10\%$. The characteristics of the family of the fundamental
solitons trapped in the isotropic and anisotropic pseudopotentials, similar
to those shown for vortices in figures \ref{fig2} and \ref{fig3}, are
displayed, respectively, in figures \ref{fig10} and \ref{fig11}.

As well as the elliptic vortices, elliptic fundamental solitons exist up to
a critical (largest) value of the eccentricity. In particular,
\begin{equation}
\left( \delta _{\mathrm{cr}}\right) _{\ell =0}=0.904  \label{cr0}
\end{equation}%
was found at $g=20$ and $g_{\mathrm{d}}=30$, cf. the similar result for the
vortices with $\ell =1$, given by equation (\ref{crit}). The critical
eccentricity is shown, as a function of the DDI strength, in figure \ref%
{fig14}, cf. figure \ref{fig6} for the vortices.
\begin{figure}[th]
\begin{center}
\includegraphics[width=0.6\columnwidth,clip]{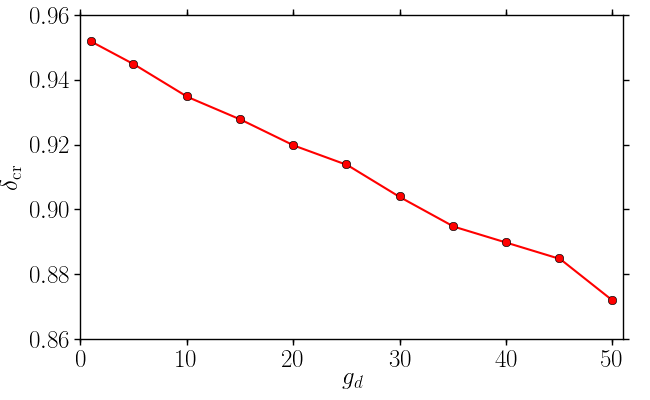}
\end{center}
\caption{(Color online) The same as in figure \protect\ref{fig6}, but for
fundamental solitons ($\ell =0$).}
\label{fig14}
\end{figure}

\begin{figure}[th]
\begin{center}
\includegraphics[width=0.9\columnwidth,clip]{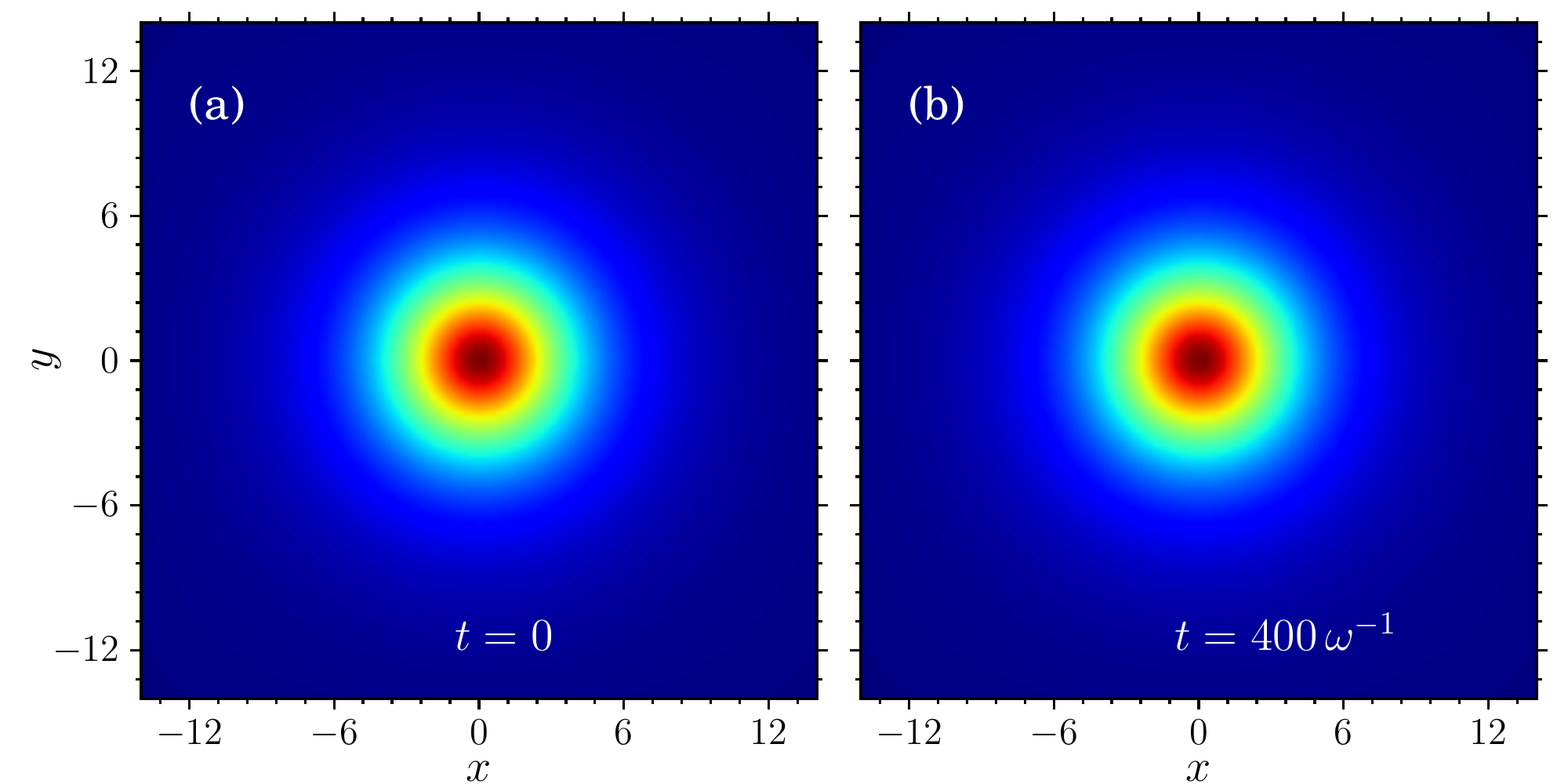}
\end{center}
\caption{(Color online) The stable evolution of a fundamental soliton, found
for the interaction strengths $g=20$, $g_{d}=30$ and isotropic
nonlinearity-modulation profile (\protect\ref{V}) with $\protect\epsilon %
_{1}=\protect\epsilon _{2}=0.1$. Shown is the initial state, produced by the
imaginary-time propagation starting from Gaussian ansatz (\protect\ref%
{eqn:vs-ansatz}) with $\ell =0$, $\protect\alpha =0.0276$ and $\protect%
\gamma =0.7203$, and the final result of subsequent real-time simulations.}
\label{fig12}
\end{figure}

\begin{figure}[th]
\begin{center}
\includegraphics[width=0.9\columnwidth,clip]{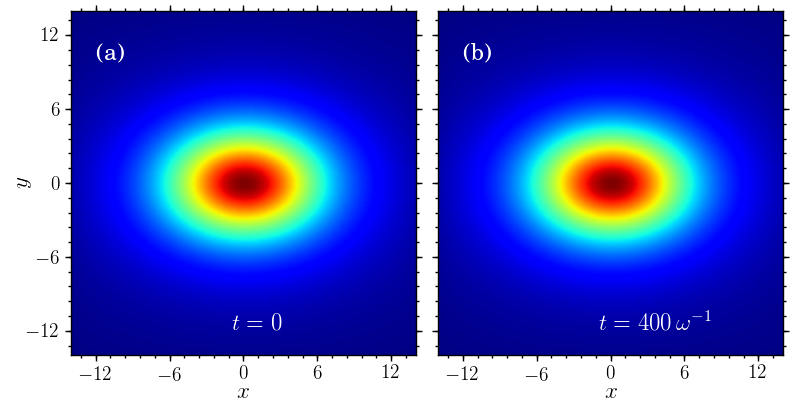}
\end{center}
\caption{(Color online) The same as in figure \protect\ref{fig12}, but for a
stable elliptic fundamental elliptic soliton trapped in the anisotropic
pseudopotential with $\protect\epsilon _{1}=0.05$, $\protect\epsilon _{2}=0.1
$. The initial state is produced by the imaginary-time propagation starting
from Gaussian ansatz (\protect\ref{eqn:vs-ansatz}) with $\protect\alpha %
=0.0229$ and $\protect\gamma =0.7657$.}
\label{fig13}
\end{figure}

Finally, the fundamental solitons, as well as their counterparts with $\ell
=1$, are stable in the isotropic and anisotropic settings alike. The
corresponding examples of the stable perturbed evolution are displayed in
figures \ref{fig12} and \ref{fig13}, respectively.

\section{Conclusion}

The objective of this work is to extend the recently introduced concept of
the self-trapping of bright solitons and solitary vortices, supported by the
repulsive local nonlinearity with the strength growing from the center to
periphery \cite{Barcelona}, to the case when this trapping mechanism
competes with the long-range repulsive DDI\ (dipole-dipole interactions) in
the dipolar condensate polarized perpendicular to the plane in which the
effective nonlinear potential provides for the self-trapping (the
confinement in the perpendicular direction is imposed by the usual linear
harmonic-oscillator trap). Another essential extension reported in this
paper is the consideration of the self-trapping of elliptic vortices and
fundamental solitons in the anisotropic nonlinear pseudopotential (\ref{V}).
By means of the VA (variational approximation) and numerical simulations in
imaginary and real time, we have constructed families of the self-trapped
vortical modes with topological charges $\ell =1$ and $\ell =2$, as well as
fundamental ones with $\ell =0$. It has been found that the fundamental
solitons and vortices with $\ell =1$ exist up to critical values of the
eccentricity [which are rather close to $1$, i.e., they correspond to the
strongly elongated elliptic shape, see equations (\ref{crit}) and (\ref{cr0}%
)]. These modes are stable in their entire existence region. On the other
hand, the double vortices with $\ell =2$ are stable only under the the
circular nonlinear pseudopotential, while the elliptic one triggers an
instability mode in the form of periodic breakdown and recovery of the
vortical mode.

A challenging possibility for developing the present analysis is to consider
a configuration with the \textit{in-plane} polarization of the dipole
moments, which will make the DDI anisotropic, cf. Ref. \cite{Malomed2008}.
An especially intriguing issue is the possibility of the existence of
vortical modes in such a setting.


\section*{Acknowledgments}

We appreciate valuable discussions with S. K. Adhikari. RKK acknowledges
support from the Third World Academy of Sciences (TWAS) and Conselho
Nacional de Desenvolvimento Cient\'{\i}fico e Tecnol\'{o}gico (CNPq, Brazil)
for the financial support in the form of a TWAS-CNPq fellowship. The work of
PM is a part of research projects funded by the Government of India, through
Department of Science and Technology (Ref. No. SR/S2/HEP-03/2009) and
Council of Scientific and Industrial Research (Ref. No. 03(1186)/10/EMR-II).
BAM\ appreciates a visitor's grant provided by the South American Institute
for Fundamental Research through Instituto de F\'{\i}sica Te\'{o}rica,
Universidade Estadual Paulista (S\~{a}o Paulo). The work of this author is
also partly supported by the German-Israel Foundation through grant No.
I-1024-2.7/2009.

\end{document}